\documentclass[aps,prb,showpacs,twocolumn]{revtex4}
\usepackage{graphicx}
\usepackage[latin1]{inputenc}
\usepackage{amsmath}
\newcommand{\eq}[1]{Eq.~(\ref{#1})}
\newcommand{\fig}[1]{Fig.~\ref{#1}}
\newcommand{\sect}[1]{Sec.~\ref{#1}}

\begin{document}
\title
{Particle number renormalization in almost half filled Mott Hubbard superconductors}
\author{Bernhard Edegger$^1$, Noboru Fukushima$^2$,
        Claudius Gros$^1$, V.N. Muthukumar$^3$}
\affiliation{$^1$ Institute for Theoretical Physics,
Universit\"at Frankfurt, D-60438 Frankfurt, Germany}
\affiliation{$^2$ Department of Physics, University of the
Saarland, D-66041 Saarbr\"ucken, Germany} \affiliation{$^3$
Department of Physics, City College of the City University of New
York, New York, NY 10031 }
%\pacs{75.30.Gw, 75.10.Jm, 78.30.-j }
\date{\today}

\begin{abstract}
The effects of the Gutzwiller projection on a BCS wave
function with varying particle number are considered. We show that a fugacity
factor has to be introduced in these wave functions when they are
Gutzwiller projected, and derive an expression for this factor
within the Gutzwiller approximation. We examine the effects of the
projection operator on BCS wave functions by calculating the
average number of particles before and after projection. We also
calculate particle number fluctuations in a projected BCS state.
Finally, we point out the differences between projecting BCS wave functions in the
canonical and grand canonical schemes, and discuss the relevance of our results for
variational Monte Carlo studies.
\end{abstract}
%%%%%%%%%%%%%%%%%%%%%
%%%%%%%%%%%%%%%%%%%%%%
\maketitle
%%%%%%%%%%%%%%%%%%%%%%%%%%%%%
\section{Introduction}
Recently, Anderson has underscored the importance of fugacity in
wave functions that do not conserve particle number \cite{pwa_tunnel}.
Following an earlier paper by Laughlin \cite{laughlin_02},
Anderson argued that a
fugacity factor should be included in variational wave
functions of the form,
\begin{equation}
P \ |\Psi_{{\rm BCS}}\rangle= P \ \prod_{k}\left(
u_k+v_k c_{k\uparrow}^\dagger c_{-k\downarrow}^\dagger
\right)|0\rangle~.
\label{Psi0_BCS}
\end{equation}
Here, $P = \prod_i (1-n_{i\uparrow} n_{i\downarrow})$ is a
(Gutzwiller) projection operator which excludes double occupancies
at sites $i$ (Ref. \onlinecite{gutzwiller_63}), and $|\Psi_{{\rm
BCS}}\rangle$, a BCS wave function. Projected wave functions of
this form were originally proposed to describe the phase diagram
of doped Mott Hubbard insulators such as the high temperature
superconductors \cite{gros_88, RMFT, atoz, ys87}. Detailed variational
Monte Carlo (VMC) studies have been carried out recently using
projected $d$-wave BCS states as variational wave functions for
the two dimensional Hubbard model, after a suitable canonical
transformation \cite{paramekanti, gros87_almost}.

Despite their simple form, projected wave functions exhibit nontrivial
properties because the projection operator acts on a quantum many body
state. The action of the projection operator (in reducing the allowed
states in the Hilbert space) concomitant with the correlations of
the quantum state being projected, leads to a variety of physical
phenomena \cite{variety}. Other interesting effects include non-trivial
matrix-element renormalization near half-filling \cite{FEMG05}
and the occurrence of superconductivity near an
(antiferromagnetic) Mott insulator \cite{afmi}.

Approximate analytical calculations with wave functions
such as \eq{Psi0_BCS} can be done using a renormalization
scheme based on the Gutzwiller approximation.
Within this approximation, the effects
of projection on the state $|\Psi_0 \rangle$ are
approximated by a classical statistical weight factor multiplying
the quantum result \cite{vollhardt_84, extensions}.
Thus, for example,
\begin{equation}
\frac
{\langle \Psi| \hat{O} |\Psi\rangle}{ \langle \Psi|\Psi\rangle}\ \approx \ g\,
\frac{
\langle \Psi_0| \hat{O} |\Psi_0\rangle}{\langle \Psi_0|\Psi_0\rangle}~,
\label{renorm}
\end{equation}
where $\hat{O}$ is any operator, and $g$, the so called Gutzwiller
factor. For example, the Gutzwiller approximation for the
kinetic energy operator
$c^\dagger_i c_j +c^\dagger_j c_i$
and the superexchange interaction between sites
$i$ and $j$, ${\vec S}_i \cdot {\vec S}_j$ yields
the Gutzwiller factors,
\begin{equation}
g_t=\frac {1-n}{1-n/2}, \quad g_s=\frac {1}{(1-n/2)^2}~,
\label{g_t_s}
\end{equation}
where $n$ is the density of electrons.
In deriving these renormalization factors,
one considers
the number of terms that contribute to
$ \langle \Psi| \hat{O} |\Psi\rangle $ and to
$\langle \Psi_0| \hat{O} |\Psi_0\rangle $ respectively.
The ratio of these two contributions is
the renormalization factor.

The renormalization factors are functions
of the \textit{local} charge density. This is a well defined quantity
when one considers for example, a projected Fermi liquid state,

\begin{equation}
P\ |\Psi_{{\rm FS}}\rangle\ =\ P\ \prod_{k<k_F}
 c_{k\uparrow}^\dagger c_{k\downarrow}^\dagger |0\rangle~.
%\label{Psi0_FS}
\end{equation}
But suppose instead, we consider wave functions such as the BCS
state in \eq{Psi0_BCS}, where the particle number fluctuates. In
this case, it is not clear what the local charge density in
\eq{g_t_s} should be. It may be argued that the correct $n$ is set
by the average particle number $\bar{N}$. But then, projecting a
BCS state changes the average particle number; \textit{i.e.}, the
average number of electrons in $|\Psi_{{\rm BCS}}\rangle $ does
not equal that in $P |\Psi_{{\rm BCS}} \rangle $. Clearly, we need
a scheme to keep track of this effect.

Note that this problem can be avoided completely, as is done in
most variational Monte Carlo (VMC) studies. Here, the particle
number is fixed (one works in a canonical ensemble), and
\eq{Psi0_BCS} replaced by,
\begin{equation}
P_N P \ |\Psi_{{\rm BCS}}\rangle= P_N P \ \prod_{k}\left(
u_k+v_k c_{k\uparrow}^\dagger c_{-k\downarrow}^\dagger
\right)|0\rangle~.
\label{pn_pg_BCS}
\end{equation}
The operator $P_N$ fixes the particle number, and the issue of
projection changing the mean particle number does not arise
\cite{gros_88}. However, there are also other VMC studies with
wave functions that do not have  fixed particle number
\cite{yokoyama_88}. Moreover, we are often interested in
carrying out analytical approximations in the spirit of
\eq{renorm}. Since such manipulations are easier done with BCS
wave functions (where the particle number is not fixed), it is
desirable to understand the effects of the projection operator on
this class of wave functions.  In this paper, we present
analytical and numerical considerations of this problem. In doing
so, we clarify the notion of fugacity introduced by Anderson
\cite{pwa_tunnel}. We also discuss the relevance
of this approach for the Gutzwiller approximation in the grand
canonical scheme and the corresponding VMC studies
\cite{yokoyama_88}.

%%%%%%%%%%%%%%%%%%%%%%%%%%%%%%%%%%%%%%%%%%%%%%%%%%%%%%%%%%%%%%%%%%%%
%Consider $|\Psi_0 \rangle$ in (\ref{pg_bcs}). The particle number of
%the state $|\Psi_0 \rangle$ is not fixed; only the average particle number $\bar{N}$
%and the standard deviation $\bar{\sigma}$ are. They are given by \cite{tinkham}
%\begin{align}
%\bar{N} &= \frac{\langle \Psi_0|\, \hat{N}
 %\,|\Psi_0 \rangle}{\langle \Psi_0|\Psi_0 \rangle} =
%2~\sum_k |v_k|^2 \label{mean_before} \\
%\bar{\sigma}^2&=\frac{\langle \Psi_0|\, (\hat{N}-\bar{N})^2
 %\,|\Psi_0 \rangle }{\langle \Psi_0|\Psi_0 \rangle}
% =4 \sum_k |v_k|^2 |u_k|^2 \label{dev_before}
%\end{align}
%%%%%%%%%%%%%%%%%%%%%%%%%%%%%%%%%%%%%%%%%%%%%%%%%%%%%%%%%%%%%%%%
\section{The fugacity factor}
\label{sec_fugacity} Consider the projected BCS wave function,
\eq{Psi0_BCS}. It is clear that the projection operator $P$ changes the average number,
\emph{viz.},
$$
{\langle \Psi_{{\rm BCS}}| \hat{N} |\Psi_{{\rm BCS}}\rangle
\over\langle \Psi_{{\rm BCS}}|\Psi_{{\rm BCS}}\rangle}
\neq
{\langle \Psi_{{\rm BCS}}|P~\hat{N}~P |\Psi_{{\rm BCS}}\rangle
\over\langle \Psi_{{\rm BCS}}|P^2 |\Psi_{{\rm BCS}}\rangle}~.
$$
The effect of the projection operator can be seen most clearly by
examining the particle number distribution in the unprojected and
projected Hilbert spaces. Towards this end, let us write the
average numbers, $\bar{N}^{(0)}(\bar{N})$ in the unprojected
(projected) Hilbert space
\begin{align}
\bar{N}^{(0)}&= \sum_N N  \, \rho^{(0)}_N~,\\
\bar{N}&= \sum_N N \, \rho_N~. \label{mean_after1}
\end{align}
Here,
\begin{eqnarray*}
\rho^{(0)}_N&=& \frac{\langle \Psi_{{\rm BCS}}|\, P_N
\,|\Psi_{{\rm BCS}}\rangle}{\langle \Psi_{{\rm BCS}}|\Psi_{{\rm BCS}}\rangle}\nonumber~, \\
{\rho}_N&=& \frac{\langle \Psi_{{\rm BCS}}|\,P~P_N~P
\,|\Psi_{{\rm BCS}}\rangle}{\langle \Psi_{{\rm BCS}}|P~P~|\Psi_{{\rm BCS}}\rangle}~,\\
\end{eqnarray*}
are the particle number distributions in the unprojected and
projected BCS wave functions respectively; $P_N$ is an operator
which projects onto terms with particle number $N$. The particle number
distributions before and after projection may be related by
\begin{align}
\underbrace{\frac{\langle \Psi_{{\rm BCS}}|\,P~P_N~P \,|\Psi_{{\rm
BCS}}\rangle}{\langle \Psi_{{\rm BCS}}|P~P~|\Psi_{{\rm
BCS}}\rangle}}_{\rho_N} = g_N \, \underbrace{\frac{\langle
\Psi_{{\rm BCS}}|\, P_N \,|\Psi_{{\rm BCS}}\rangle}{\langle
\Psi_{{\rm BCS}}|\Psi_{{\rm BCS}}\rangle}}_{\rho^{(0)}_N} \,
\label{gutzP} ,
\end{align}
where
\begin{align}
g_N=\underbrace{\frac{\langle \Psi_{{\rm BCS}}|\Psi_{{\rm
BCS}}\rangle}{\langle \Psi_{{\rm BCS}}|P~P~|\Psi_{{\rm
BCS}}\rangle}}_{=C(=\rm const)} \frac{\langle \Psi_{{\rm BCS}}|\,
P~P_N~P \,|\Psi_{{\rm BCS}}\rangle}{\langle \Psi_{{\rm
BCS}}|\,P_N\,|\Psi_{{\rm BCS}}\rangle} \nonumber \ .
\end{align}
\eq{gutzP} constitutes the Gutzwiller approximation for the
projection operator $P_N$ with the corresponding renormalization
factor, $g_N$; $C$ is an irrelevant constant (the ratio of the
normalization of the unprojected and projected wave functions),
which does not depend on $N$. Following Gutzwiller, we estimate
$g_N$ by combinatorial means, as being equal to
the ratio of the relative sizes of the projected and unprojected
Hilbert spaces. Then,
\begin{align}
g_N\approx C \, \frac{\frac{L!}{(L-N_{\uparrow}-N_{\downarrow})!
\,N_{\uparrow}!\,N_{\downarrow}!}}
{\frac{L!}{(L-N_{\uparrow})!\,N_{\uparrow}!}\,
\frac{L!}{(L-N_{\downarrow})!\,N_{\downarrow}!}}\, ,
\end{align}
where $L$ is the number of lattice sites and $N_{\uparrow}$
($N_{\downarrow}$) is the number of up (down)-spins. Since in a
BCS wave function, $N_{\uparrow}=N_{\downarrow}=N/2$, $N$ being
the total number of particles, the expression for $g_N$ can be
simplified to
\begin{align}
g_N\approx C \, \frac{(\left (L-N/2)! \right)^2}{L!\,(L-N)!}\, .
\label{gN}
\end{align}

Hence, if we were to impose the condition that the average
particle number before and after projection be identical, a factor
$g_N^{-1}$ has to be included in \eq{mean_after1}. Then, from
\eq{mean_after1} and \eq{gutzP}, we obtain the particle number
after projection $\bar{N}_{{\rm new}}$,
\begin{align}
\bar{N}_{{\rm new}}&\equiv\   \sum_N N\,\frac{1}{g_N}\, \rho_N
\ =\  \sum_N\, N \, \frac{g_N\,\rho^{(0)}_N}{g_N} \nonumber \\
&=\ \bar{N}^{(0)} ~, \label{mean_after2}
\end{align}
which is the desired result.

Now, let us show how this
procedure can be implemented for the wave function
$|\Psi_{{\rm BCS}}\rangle$.
Since the BCS wave function is a linear superposition of states with
particle number $\ldots, N-2, N, N+2, \ldots $, we consider the
effect of projection on two states whose particle numbers differ by $2$.
Then, the ratio,
\begin{equation}
f^2 \ \equiv\  \frac{g_{N+2}}{g_N}
\ \approx\ \left(\frac{L-N}{L-N/2}\right)^2\, \label{define_f},
\end{equation}
in the thermodynamic limit. Eq.~\ref{define_f} shows that the
projection operator acts unequally on the $N$ and $N+2$ particle
states; the renormalization of the weight of the $N+2$ particle
states ${g_{N+2}}$, is ${f^2}$ times the
weight of the $N$ particle states, ${g_N}$. This effect
can be rectified as in \eq{mean_after2} by multiplying every
Cooper pair $c_{k\uparrow}^\dagger c_{-k\downarrow}^\dagger$ by a
factor $\frac{1}{f}$ in the BCS wave function. It produces the
desired result, \textit{viz.}, the projected and unprojected BCS
wave functions have the same average particle number.

Alternatively (following Anderson),
we can multiply every empty state by the factor $f$ and write,
\begin{align}
|{\Psi}^{(f)}_{{\rm BCS}}\rangle =
 \prod_k \frac{\left(f\,u_k + v_k c_{k \uparrow}^
\dagger c_{-k \downarrow}^\dagger \right)}
{\sqrt{f^2 |u_k|^2 +|v_k|^2}}\,| 0 \rangle ~.\label{new_BCS}
\end{align}
Then again by construction, the fugacity factor $f$ in
\eq{new_BCS} ensures that the projected wave function
$P|\Psi^{(f)}_{{\rm BCS}}\rangle$ and the unprojected wave
function $|\Psi_{{\rm BCS}}\rangle$ have the same particle number.
The denominator in \eq{new_BCS} is the new normalization factor.

The following points are in order: (a) the fugacity factor $f$ in
\eq{define_f} depends on the variable particle number $N$.
However, since the particle number of the BCS wave function is
sharply peaked within the range,
$\bar{N}^{(0)}-\sqrt{\bar{N}^{(0)}}$ and
$\bar{N}^{(0)}+\sqrt{\bar{N}^{(0)}}$, we will assume that the
fugacity factor $f=f(\bar{N}^{(0)})$ in the thermodynamic limit;
(b) in this limit, \eq{define_f} reduces to $f^2 = g_t^2$, where
$g_t$ is the Gutzwiller factor defined in \eq{g_t_s}. Then,
\eq{new_BCS} reduces to
\begin{align}
|{\Psi}^{(f)}_{{\rm BCS}}\rangle =
 \prod_k \frac{\left(g_t\,u_k + v_k c_{k \uparrow}^
\dagger c_{-k \downarrow}^\dagger \right)}
{\sqrt{g_t^2 |u_k|^2 +|v_k|^2}}\,| 0 \rangle ~,\label{pwa_BCS}
\end{align}
which is the wave function proposed by Anderson \cite{pwa_tunnel};
(c) the fugacity factor ensures that projection affects the $N$
and $N+2$ particle states of the BCS wave function in the same
way. In principle, such a factor could depend on the $k$-value,
but in this paper, we will treat it as a mere combinatorial device; (d) the
combinatorial argument fails at half filling when $L=\bar
N^{(0)}$.

\section{Particle number renormalization
in projected BCS wave functions}

In the previous section, we showed that the inclusion of the
fugacity factor is necessary for the average particle number in a
BCS wave function to be unchanged by projection. Alternatively,
one might ask what is the effect of the projection operator on a
BCS wave function; \textit{viz.}, if projection changes the mean
particle number of a BCS state, how are the particle numbers
before and after projection related? In this section, we will use
the fugacity factor to answer this question. In particular, we
will show how particle density after projection can be determined
self consistently by including the fugacity factor in the projected BCS
state (\eq{pwa_BCS}). An alternate derivation of this result is
presented in Appendix A, where the effect of the
projection operator on particle number fluctuations is calculated.

Consider two BCS states defined by
\begin{align}
|\Psi_{{\rm BCS}}\rangle &=
 \prod_k \left(u_k + v_k c_{k \uparrow}^
\dagger c_{-k \downarrow}^\dagger \right)\,|0\rangle \ , \\
|{\Psi}^{(r)}_{{\rm BCS}}\rangle &=
 \prod_k \frac{\left(u_k + g_t \, v_k \, c_{k \uparrow}^
\dagger c_{-k \downarrow}^\dagger \right)}
{\sqrt{|u_k|^2 +g_t^2 |v_k|^2}}\,| 0 \rangle \nonumber \\
&= \prod_k \left({u}^{(r)}_k + {v}^{(r)}_k c_{k \uparrow}^ \dagger
c_{-k \downarrow}^\dagger \right)\,| 0 \rangle \, , \label{fug_remove_state}
\end{align}
where,
\begin{align}
{u}^{(r)}_k&\equiv \frac{u_k}{\sqrt{ | u_k|^2 +g_t^2 | v_k|^2}} \label{uk}, \\
{v}^{(r)}_k&\equiv \frac{g_t \, v_k}{\sqrt{ |u_k|^2 + g_t^2 | v_k|^2}}
\label{vk}.
\end{align}

From \eq{define_f}, it is clear that the projection operator reduces the
ratio of the weights of $N+2$ and $N$ particle states in a BCS
wave function by a factor $g_t$. Then, it follows that
\begin{align}
 \frac{\langle \Psi^{(r)}_{{\rm BCS}}|\, \hat{N}
 \,|\Psi^{(r)}_{{\rm BCS}} \rangle}
 {\langle \Psi^{(r)}_{{\rm BCS}}|\Psi^{(r)}_{{\rm
 BCS}}\rangle}\approx
 \frac{\langle {\Psi}_{{\rm BCS}}|\,P\,\hat{N}\,P
 \,|{\Psi}_{{\rm BCS}}\rangle}
{\langle {\Psi}_{{\rm BCS}}|\,P\,|{\Psi}_{{\rm BCS}}\rangle}
\label{r1}
\end{align}
when
\begin{align}
g_t=\frac{L-\bar{N}^{(r)}}{L-\bar{N}^{(r)}/2}~.
\end{align}
The average particle number
$\bar{N}^{(r)}$ of the state $|\Psi^{(r)}_{{\rm BCS}} \rangle$ is
given by,
\begin{equation}
\bar{N}^{(r)} \ =\
\frac{\langle \Psi^{(r)}_{{\rm BCS}}|\,\hat{N}
 \,|\Psi^{(r)}_{{\rm BCS}}\rangle}
{\langle \Psi^{(r)}_{{\rm BCS}}|\Psi^{(r)}_{{\rm BCS}}\rangle}
 \ =\  2\,\sum_k |v_k^{(r)}|^2 \label{p_after1}.
\end{equation}
Since the particle numbers of $|\Psi^{(r)}_{{\rm BCS}}\rangle$ and
$P|\Psi_{{\rm BCS}}\rangle $ are identical, we can use \eq{vk} in
\eq{p_after1} to obtain,
\begin{align}
\bar{N}^{(r)} \approx \bar N_{\rm after} &=  \frac{\langle
{\Psi}_{{\rm BCS}}|\,P\,\hat{N}\,P
 \,|{\Psi}_{{\rm BCS}}\rangle}
{\langle {\Psi}_{{\rm BCS}}|\,P\,|{\Psi}_{{\rm
BCS}}\rangle}
\nonumber \\
&\approx 2\,\sum_k \frac{g_t^2 |{v}_k|^2} {|{u}_k|^2 +g_t^2
|{v}_k|^2} \label{p_after2}~ .
\end{align}
Note that $g_t$ is specified by the particle number after
projection, $N_{\rm after}\,(=\bar{N}^{(r)})$.

Now, since the number of particles in the state $|{\Psi}_{{\rm
BCS}} \rangle$ before projection is given by
$$
\bar N_{\rm before} = 2\, \,\sum_k |{v}_k|^2~,
$$
\eq{p_after2} provides us with a way to calculate the number of
particles in the state $P\,|{\Psi}_{{\rm BCS}} \rangle$
\emph{after} projection, if $|{\Psi}_{{\rm BCS}}\rangle$
(\textit{i.e.}, $u_k$ and $v_k$) is specified \textit{before}
projection. \eq{p_after2} can be solved self consistently for
$\bar{N}_{{\rm after}}$. We solve \eq{p_after2} numerically on a square
lattice, using the standard BCS expressions for a $d$-wave
superconductor, $u_k$($v_k$):
\begin{align}
v^2_k&=\frac{1}{2} \left( 1 - \frac{\xi_k}{E_k}  \right)~,\label{start_dwave}\\
u^2_k&=\frac{1}{2} \left( 1 + \frac{\xi_k}{E_k}  \right)~,
\end{align}
where,
\begin{align}
E_k &=\left( \Delta_k^2 + \xi_k^2  \right)^{\frac 1 2},\\
\Delta_k &=\Delta_0\,(\cos(k_x)-\cos(k_y)),\\
\xi_k &=-2\,(\cos(k_x)+\cos(k_y))-\mu~. \label{end_dwave}
\end{align}
The only free parameters are the chemical potential $\mu$ and the variational
parameter $\Delta_0$.

\begin{figure}[t]
  \centering
 \includegraphics*[width=0.45\textwidth]{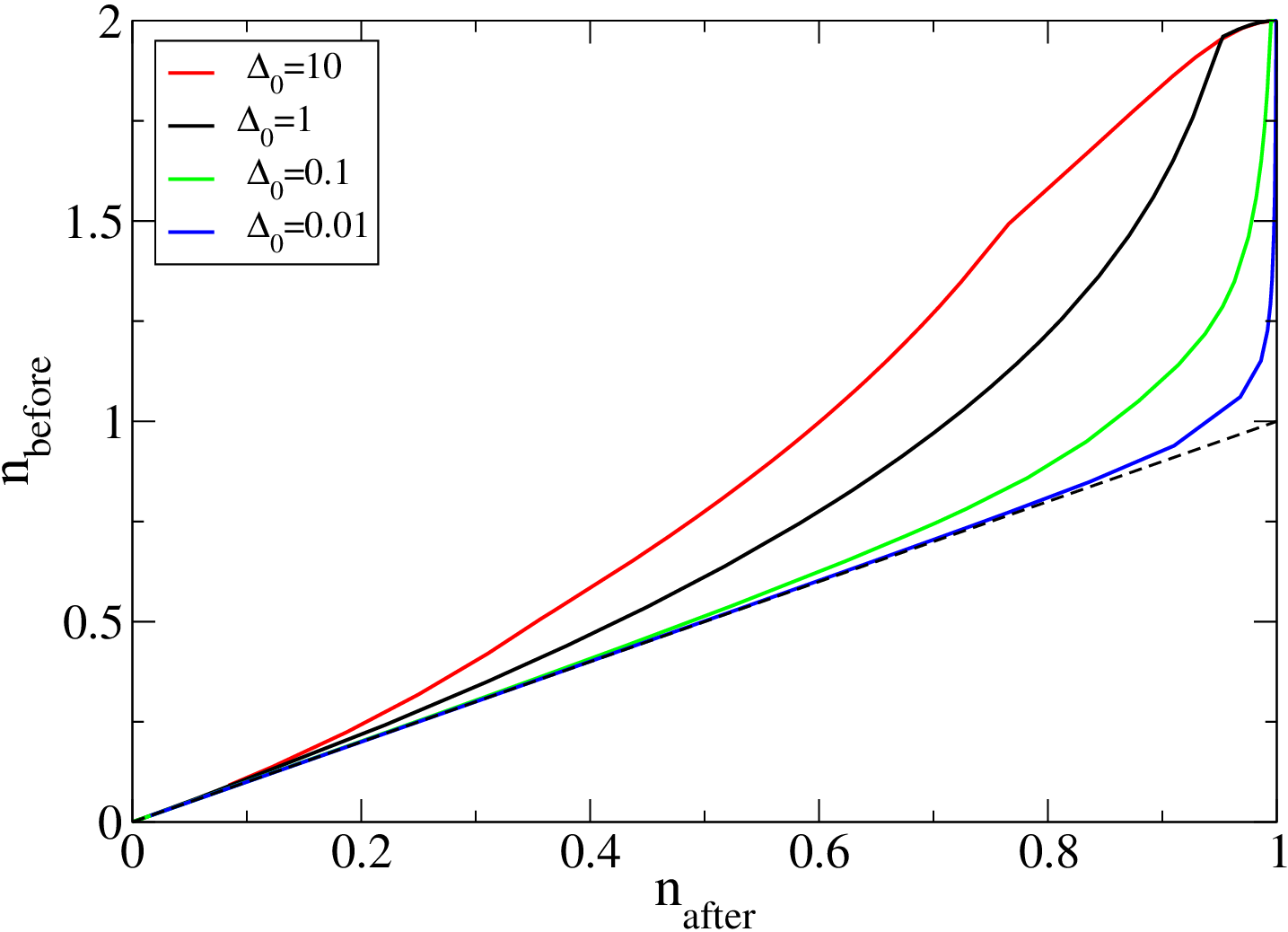}
 \caption{
(Color online)
Particle density before projection $n_{{\rm before}}$ as a function
of the particle density after projection $n_{\rm after}$ for
$d$-wave states with different parameters $\Delta_0$.
The dashed line indicates the Fermi liquid
result $n_{\rm before}=n_{\rm after}$. \label{fig:nf_na}}%
\end{figure}

By fixing the parameter $\Delta_0$, we determine the
particle numbers (before and after projection) for various
chemical potentials. The results for particle density ($n \equiv
\bar N/L$) are shown in \fig{fig:nf_na}. The results clearly show that
the particle density before projection attains its maximal
value ($n_{\rm before}=2$), if $n_{\rm after}=1$ (half-filling). This
result holds for any finite value of the variational parameter
$\Delta_0$. In the opposite limit, \textit{viz.}, low density of
electrons, $n_{\rm before}$ converges to the value of $n_{\rm after}$ as
expected. The size of the intermediate region depends on the
magnitude of the parameter $\Delta_0$, as illustrated by the results in
\fig{fig:nf_na}.

The accuracy of \eq{p_after2} can be checked by comparing our
results with those of Yokoyama and Shiba (YS), who performed VMC
studies of projected BCS wave functions with fluctuating particle
number (but without the fugacity factor) \cite{yokoyama_88}. They
determined the particle density of the projected $d$-wave state
$P\,|\Psi_{{\rm BCS}} \rangle$ as a function of the chemical
potential $\mu$ and the variational parameter $\Delta_0$, within a grand
canonical scheme. The unprojected wave function $|\Psi_{{\rm BCS}}
\rangle$ is specified as usual, through
\eq{start_dwave}-\eq{end_dwave}. Since YS do not include a
fugacity factor in their definition of the BCS wave function,
projection changes the particle number. So, we use \eq{p_after2}
to determine $n_{\rm after}$ which we compare with their results for
particle number.

\begin{figure}
  \centering
  \includegraphics*[width=0.45\textwidth]{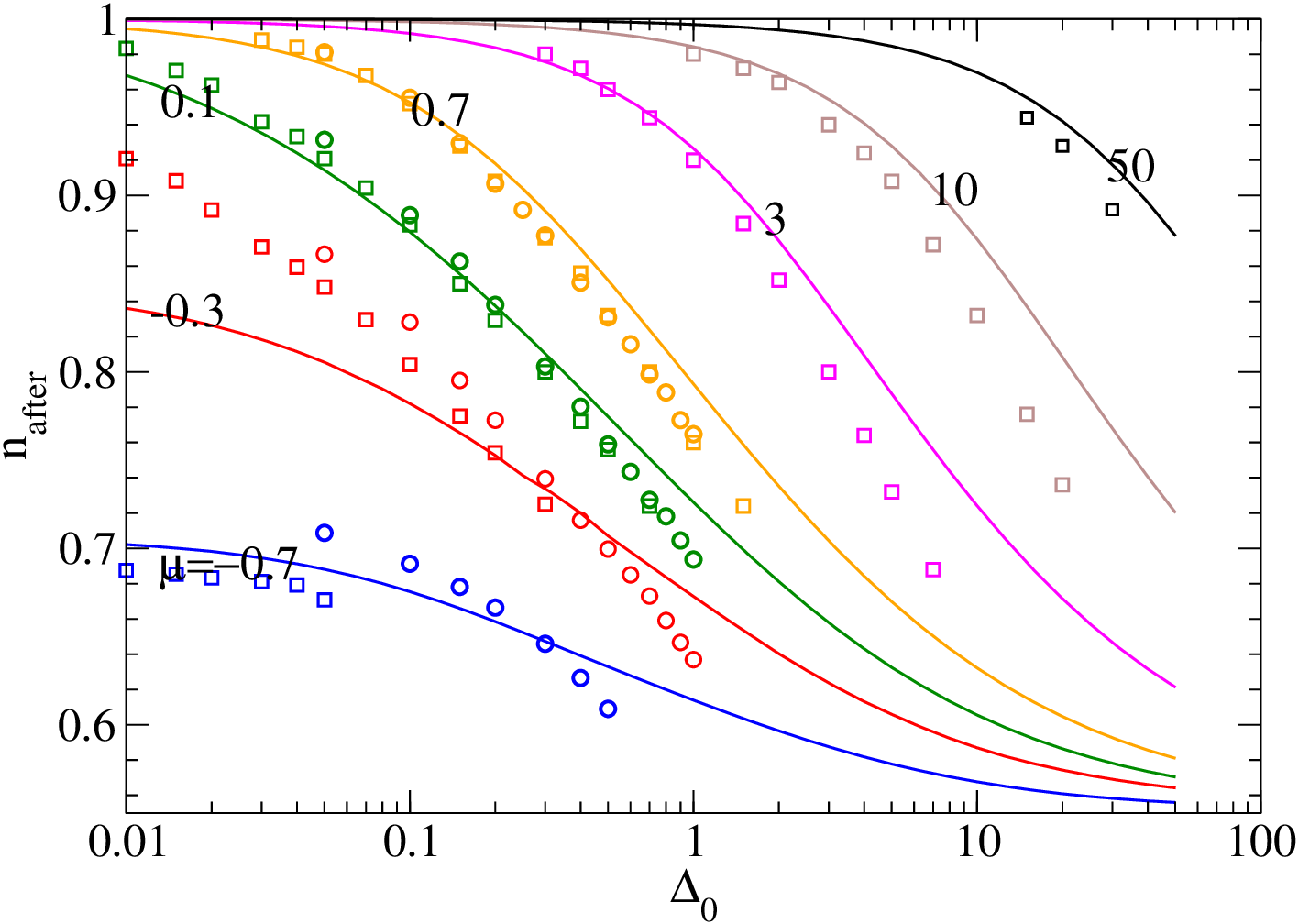}
  \caption{\label{fig:compare_Yoko}%
(Color online)
The particle density  after projection $n_{\rm after}$ as a
function of parameter $\Delta_0$ for a $d$-wave BCS
state at various chemical potentials $\mu$. The figure shows a
comparison between results from \eq{p_after2}(solid lines) and the
VMC results of Yokoyama and Shiba \cite{yokoyama_88} (for
$6\times6$ - (circles) and $8\times8$ -lattices (squares)).
Numbers in the figure denote the chemical potentials of the
corresponding curves and data points.}
\end{figure}

As seen in \fig{fig:compare_Yoko}, our results
are in good qualitative agreement with YS. Discrepancies
are mostly due to finite corrections. YS use $6\times6$
and $8\times8$-lattices, while our analytic calculations
are for the thermodynamic limit\cite{infiniteS}.
The results show the singular effect of the projection near the
insulating phase (half filling). The chemical potential goes to
infinity in this limit.

%\begin{figure}
%  \centering
%  \includegraphics*[width=0.45\textwidth]{ne_d0.eps}
%  \caption{\label{fig:ne_d0}
%The particle density after projection $n_{\rm after}$ as a function of
%the SC-order parameter $\Delta_0$ for a $d$-wave BCS state at various
%chemical potentials $\mu$. The results are determined by solving
%\eq{p_after2} self consistently. Numbers in the figure denote the chemical
%potentials of the corresponding curves.
%}
%\end{figure}

In Appendix A, we present an alternate derivation of \eq{p_after2}
by a saddle point approximation
without using a fugacity-corrected wave function. This approach
also allows for the calculation of particle number fluctuations $
{\sigma^2}/L$.
%%%%%%%%%%%%%%%%%%%%%%%%%%%%%%%%%%%%%%%%%%%%%%%%%%%%%%%%%%%%
\section{Gutzwiller approximation in the canonical and grand canonical schemes}

In this section, we discuss the differences between the Gutzwiller
approximation in the canonical and grand canonical schemes. The
validity of our statements can be checked by a comparison to nearly
exact VMC \cite{yokoyama_88,gros_88,paramekanti,yokoyama_96}.

Let us first consider the canonical case. Here, we are
interested in the expectation value of an operator $\hat O$
calculated with a particle number conserving projected wave function
$P_N P \, |\,\Psi_{{\rm BCS}}\rangle$. The corresponding
Gutzwiller approximation can be understood as follows:
\begin{align}
&\frac
{\langle \Psi_{{\rm BCS}}|\, P~P_N~\hat{O}~P_N~P \, |\,\Psi_{{\rm BCS}}\rangle}{\langle
  \Psi_{{\rm BCS}}|\,  P~P_N~P\, |\,\Psi_{{\rm BCS}}\rangle} \nonumber \\
\approx & \ g\,
\frac
{\langle \Psi_{{\rm BCS}}|\, P_N~\hat{O}~P_N \,|\,\Psi_{{\rm BCS}}\rangle}{\langle
  \Psi_{{\rm BCS}}| \,P_N \,|\,\Psi_{{\rm BCS}}\rangle} \nonumber \\
= & \ g\,
\frac
{\langle \Psi_{{\rm BCS}}|\, \hat{O}\, |\,\Psi_{{\rm BCS}}\rangle}{\langle
  \Psi_{{\rm BCS}}|\,\Psi_{{\rm BCS}}\rangle}\  ,
\label{fixedN}
\end{align}
where $P_N$ is the projector on the terms with particle number
$N$. The Gutzwiller factor $g$, corresponds to the
operator $\hat O$. The first row represents a quantity which can be
calculated exactly by fixed particle number VMC
\cite{gros_88,paramekanti}. Since the particle number is fixed,
the Gutzwiller approximation can be invoked, leading to the second
row. The equality to the third row is guaranteed only if $N$ is
equal to the average particle number of ${|\,\Psi_{{\rm
BCS}}\rangle}$ ($N=\bar N$). Here, we perform a transformation
from a canonical to a grand canonical ensemble, which is
valid in the thermodynamic limit.

In the grand canonical scheme, where we calculate the
expectation value of $\hat O$ with a particle number non-conserving
wave function, this scheme must be modified as follows:
\begin{equation}
\frac
{\langle \Psi^{(f)}_{{\rm BCS}}|\, P~\hat{O}~P\, |\,\Psi^{(f)}_{{\rm BCS}}\rangle}{\langle
  \Psi^{(f)}_{{\rm BCS}}|\, P~P \,|\, \Psi^{(f)}_{{\rm BCS}}\rangle}
\ \approx\   \ g\,
\frac
{\langle \Psi_{{\rm BCS}}|\, \hat{O} \,|\,\Psi_{{\rm BCS}}\rangle}{\langle
  \Psi_{{\rm BCS}}|\,\Psi_{{\rm BCS}}\rangle}\  ,
\label{grandC}
\end{equation}
where ${P |\, \Psi^{(f)}_{{\rm BCS}}\rangle}$ is the projected
$d$-wave state corrected for fugacity, \textit{i.e.}, a fugacity
factor is included simultaneously with the projection (see
\sect{sec_fugacity}). This correction is essential to guarantee
the validity of the Gutzwiller approximation; without it,
the lhs  and rhs of \eq{grandC} would correspond to
states with different particle numbers.

Comparing \eq{fixedN} and \eq{grandC} we get,
\begin{equation}
\frac{\langle \Psi_{{\rm BCS}}| P P_N \hat{O} P_N P |\,\Psi_{{\rm
BCS}}\rangle}{\langle
  \Psi_{{\rm BCS}}|  P P_N P|\,\Psi_{{\rm BCS}}\rangle}
\ \approx \ \frac{\langle \Psi^{(f)}_{{\rm BCS}}| P \hat{O}
P|\,\Psi^{(f)}_{{\rm BCS}}\rangle}{\langle
  \Psi^{(f)}_{{\rm BCS}}| P^2 |\, \Psi^{(f)}_{{\rm BCS}}\rangle}\ . \label{comp_eq}
\end{equation}
\eq{grandC} and \eq{comp_eq} constitute the main results of this section.
\eq{grandC} shows that when the Gutzwiller approximation is used for a wave function
which does not have a fixed particle number, a fugacity factor must be included along with the projection. \eq{comp_eq} shows that to obtain
identical results, one has to use
different wave functions
in the grand canonical (rhs) and
canonical (lhs) schemes. The wave function
${|\,\Psi^{(f)}_{{\rm BCS}}\rangle}$ is a $d$-wave state corrected
by the fugacity factor, whereas ${|\,\Psi_{{\rm BCS}}\rangle}$ is
a pure $d$-wave state. Our arguments leading up to \eq{grandC} and \eq{comp_eq} can be verified by a comparison with VMC studies. We now proceed to do so.

\begin{figure}
  \centering
  \includegraphics[width=0.45\textwidth]{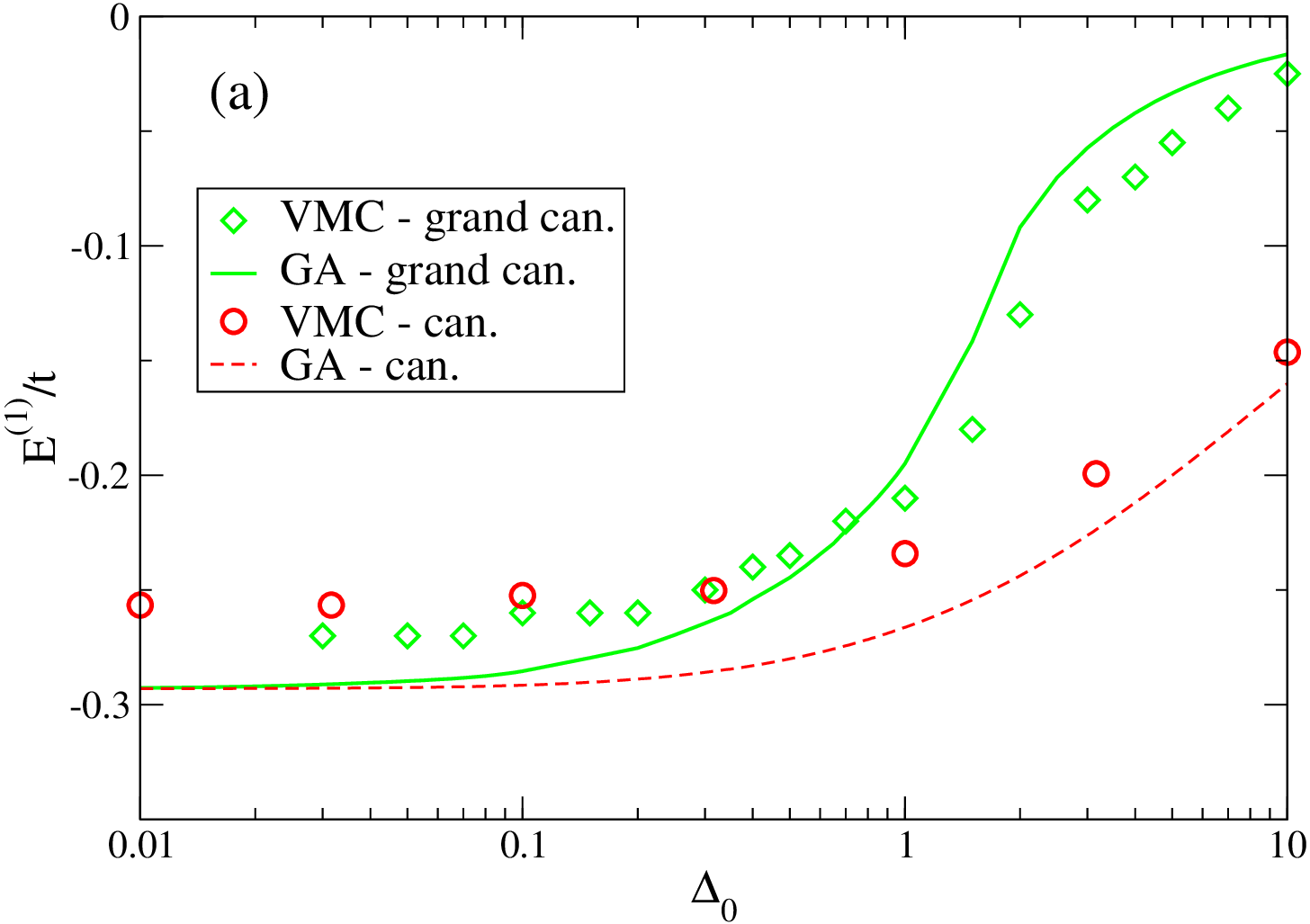}
  \includegraphics[width=0.45\textwidth]{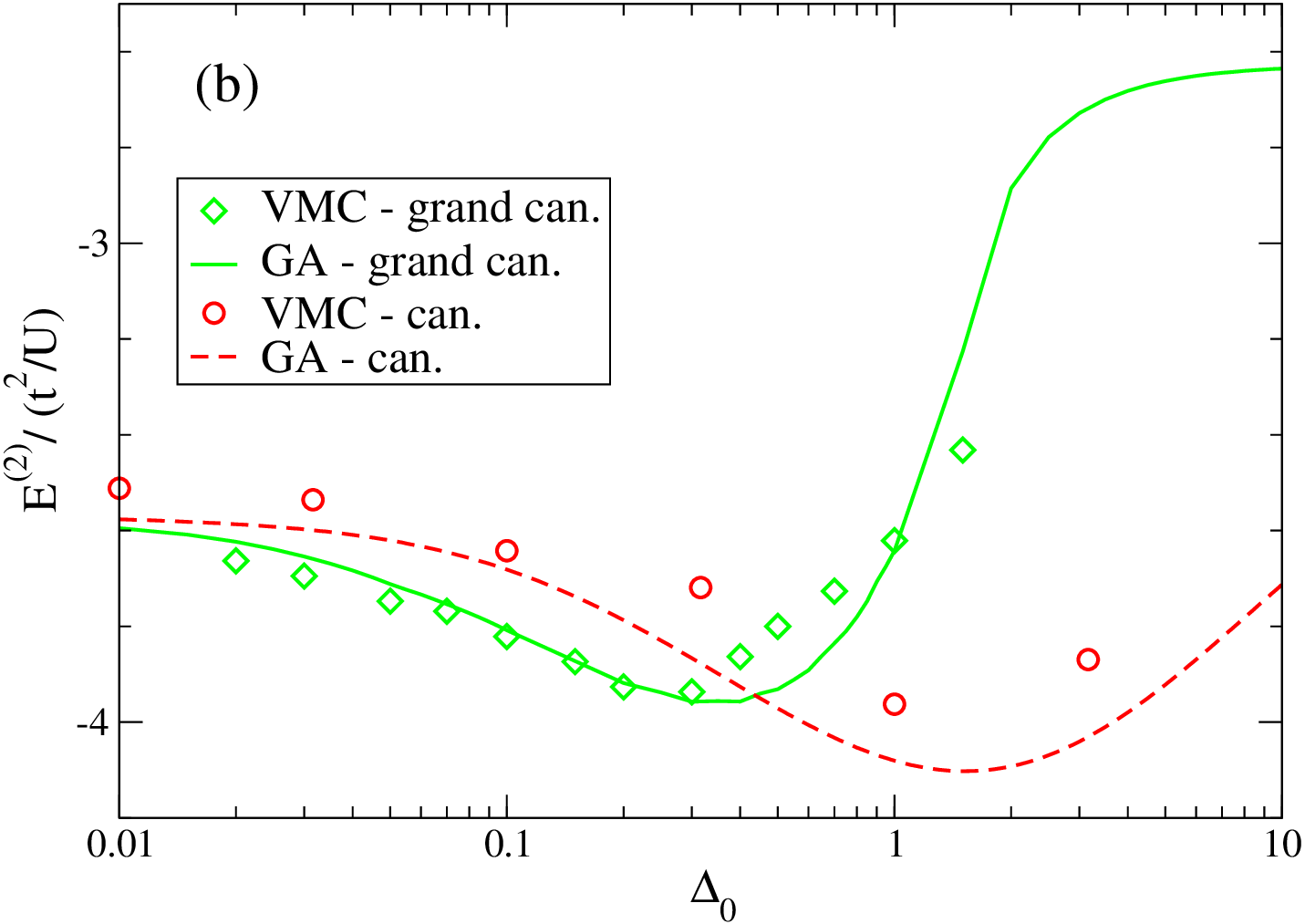}
  \caption{
(Color online)
(a) The kinetic energy $E^{(1)}$
and (b) the energy of the remaining terms $E^{(2)}$ per site of
the $t-J$ model as a function of the variational
parameter $\Delta_0$ for the $d$-wave state
at a filling $n=0.9$. Fixed particle (can.) VMC
data\cite{gros_88} (circles, 82 sites) and grand canonical
VMC\cite{yokoyama_88} (squares, $8\times8$ sites) are compared.
The dashed/solid lines represent the corresponding Gutzwiller
approximations (GA). For a detailed description, we refer to the
text and Appendix B.} \label{compare_VMC}
\end{figure}

The expectation values in the canonical and grand canonical schemes
can be calculated (nearly exactly) by VMC studies. In
\fig{compare_VMC}, we show VMC results from Gros \cite{gros_88}
(fixed particle number VMC, canonical) and from YS
\cite{yokoyama_88} (grand canonical VMC). The discrepancy between
the two sets of results can be explained readily by \eq{comp_eq}.
In the case of $\Delta_0 \rightarrow 0$, there is only small room
for particle number fluctuations even in the particle
non-conserving wavefunction. Then, canonical and grand canonical
schemes should give identical results. The VMC calculations in
\fig{compare_VMC} do not exactly show thit behavior since the
grand-canonical scheme becomes inaccurate in this limit
\cite{yokoyama_88}. YS consider a pure $d$-wave state,
\textit{i.e.}, the fugacity factor is not included in their
calculations. In their paper, YS argued that the discrepancies
between the two results can be removed by introducing an
additional variational parameter $\alpha$, so that $a_k \equiv
v_k/u_k$ is replaced by $a_k \equiv \alpha \, v_k/u_k$ (Eq. 4.1 in
Ref. \onlinecite{yokoyama_88}). We opine that the parameter
$\alpha$ is directly related to our fugacity factor,
\textit{i.e.}, $\alpha=g_t$ in the wave function
${|\,\Psi^{(f)}_{{\rm BCS}}\rangle}$. This conclusion is supported
by the comparison of VMC data to the corresponding Gutzwiller
approximation (see below).

The validity of the approximation in the canonical case
(\eq{fixedN}) is well accepted. It is used for instance, in the
renormalized mean field theory (RMFT) of Zhang \emph{et al.},
where all physical quantities are calculated using unprojected
wave functions and the corresponding Gutzwiller renormalization
factors \cite{RMFT}. A comparison with VMC studies with fixed
particle number shows good agreement \cite{RMFT} (also
illustrated in \fig{compare_VMC}).

To compare the grand canonical VMC of YS with the Gutzwiller
approximation, we need to modify \eq{grandC}. This is necessary because YS do not include the fugacity factor in their considerations, as pointed out earlier.
%  We remove the fugacity factor $g_t$ on both sides of \eq{grandC} and write,
%  \begin{align}
%  &\frac
%  {\langle \Psi_{{\rm BCS}}|\, P~\hat{O}~P\, |\, \Psi_{{\rm BCS}}\rangle}{\langle
%     \Psi_{{\rm BCS}}|\, P~P \,|\,  \Psi_{{\rm BCS}}\rangle} \nonumber \\
%  \approx & \ g\,
%  \frac
%  {\langle  \Psi^{(r)}_{{\rm BCS}}|\, \hat{O} \,|\, \Psi^{(r)}_{{\rm BCS}}\rangle}{\langle
%   \Psi^{(r)}_{{\rm BCS}}|\,\Psi^{(r)}_{{\rm BCS}}\rangle}\  ,
%  \label{grandC2}
%  \end{align}
%  where ${|\,\Psi^{(r)}_{{\rm BCS}}\rangle}$ is a ``fugacity factor removed''
%  wave function, \textit{i.e.}, ${|\,\Psi^{(r)}_{{\rm BCS}}\rangle}$ is given as in
%  \eq{fug_remove_state}, where $u_k (v_k)$
%  satisfy the $d$-wave definitions. Note the analogy with \eq{r1}. Now, the Gutzwiller
%  approximation in \eq{grandC2} (second row) corresponds
%  to a grand canonical VMC with an uncorrected $d$-wave function
%  ${|\,\Psi_{{\rm BCS}}\rangle}$ as in YS.
%  \textbf{How is $g$ specified in \eq{grandC2}? By $\bar{N}^{(r)}$, right?}
We modify \eq{comp_eq}  by the following procedure:\newline (i) we start with a
$d$-wave BCS state ${|\, \Psi_{{\rm BCS}}\rangle}$ for specified
values of $\Delta_0$; \newline %
(ii) we use \eq{p_after2} to
determine the chemical potential $\mu$. This fixes the particle
density $n_{\rm after}$ of ${P|\, \Psi_{{\rm BCS}}\rangle}$; \newline
(iii) we remove the fugacity factor to get ${|\,\Psi^{(r)}_{{\rm
BCS}}\rangle}$ via Eq.\ (\ref{fug_remove_state}). The fugacity
factor is determined for $n_{\rm after}$. ${|\,\Psi^{(r)}_{{\rm
BCS}}\rangle}$ and ${P|\, \Psi_{{\rm BCS}}\rangle}$ correspond to
the same particle density $n_{\rm after}$.
\newline%
(iv) The expectation values of the wave function ${P|\, \Psi_{{\rm
BCS}}\rangle}$ can now be approximated by ${|\, \Psi^{(r)}_{{\rm
BCS}}\rangle}$ and Gutzwiller factors, {\em viz.},
\begin{equation}
\frac
{\langle \Psi_{{\rm BCS}}|\, P~\hat{O}~P\, |\, \Psi_{{\rm BCS}}\rangle}{\langle
  \Psi_{{\rm BCS}}|\, P~P \,|\,  \Psi_{{\rm BCS}}\rangle}
\ \approx \ g\,
 \frac
{\langle  \Psi^{(r)}_{{\rm BCS}}|\, \hat{O} \,|\, \Psi^{(r)}_{{\rm BCS}}\rangle}{\langle
  \Psi^{(r)}_{{\rm BCS}}|\,\Psi^{(r)}_{{\rm BCS}}\rangle}\  .
\label{grandC2}
\end{equation}

This Gutzwiller approximation (GA) generalizes Eq.\ \ref{renorm}
for wave functions that do not conserve particle number.
In Appendix B, we discuss this approximation for the different terms in the $t-J$ model.

\begin{figure}
  \centering
  \includegraphics[width=0.45\textwidth]{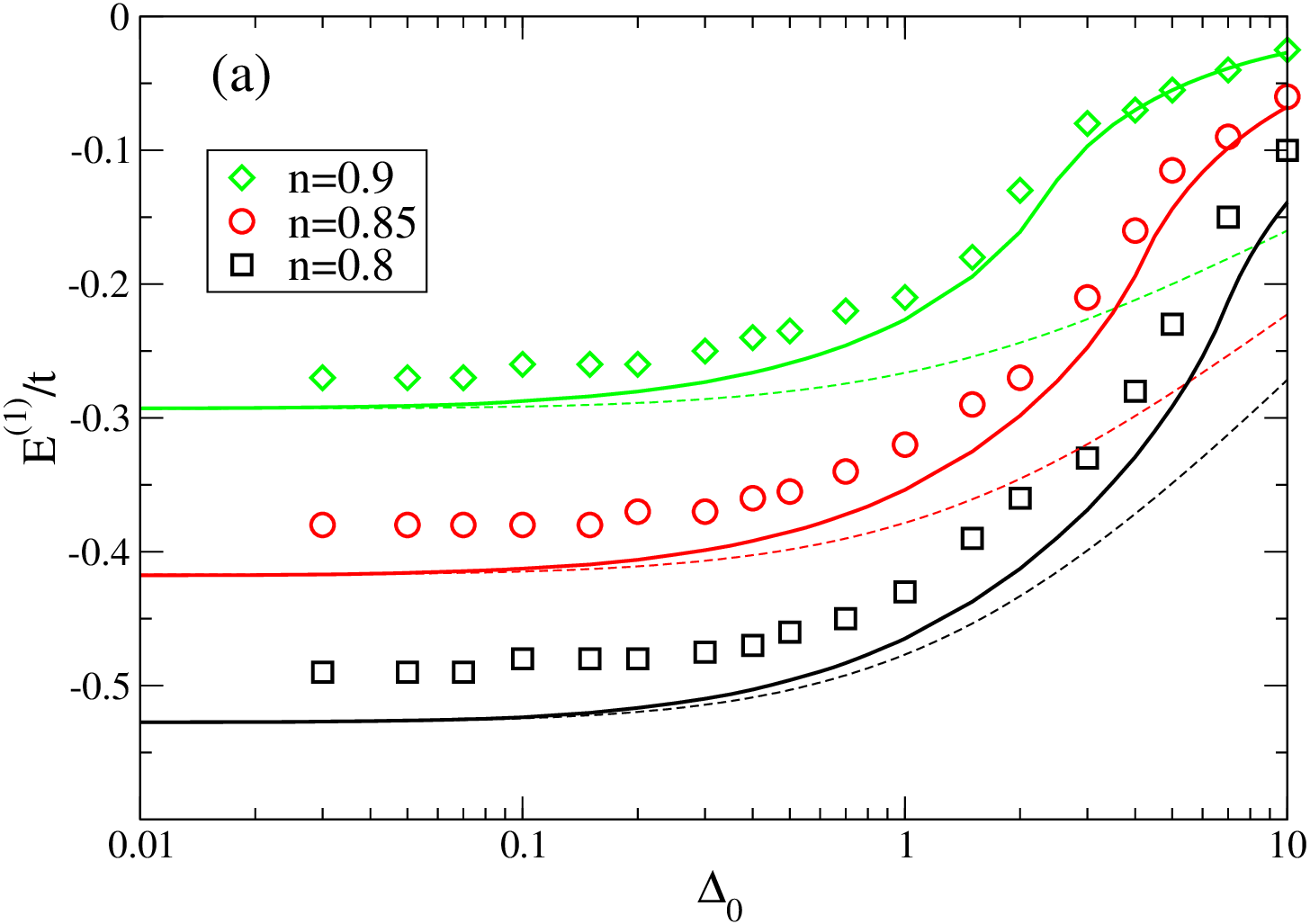}
  \includegraphics[width=0.45\textwidth]{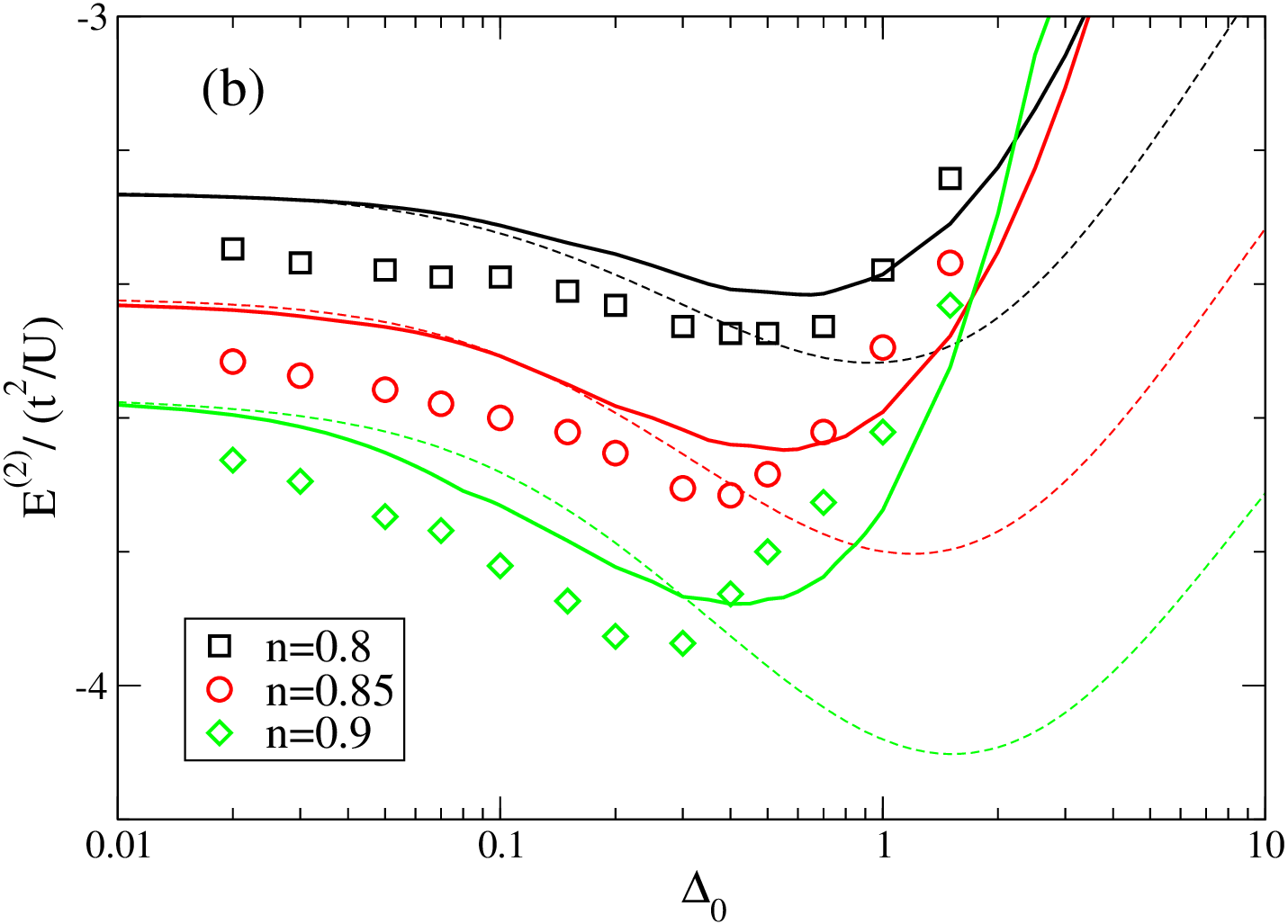}
  \caption{
(Color online)
(a) The kinetic energy $E^{(1)}$ and (b) the energy of the remaining
terms $E^{(2)}$ per site of the $t-J$ model, as a function of the variational
parameter $\Delta_0$ for the $d$-wave state at various densities. The data
points (diamond, circle, square) are taken from the grand
canonical VMC study of Yokoyama and Shiba \cite{yokoyama_88}. The
solid/dashed lines represent
the corresponding Gutzwiller approximations for a
grand canonical/fixed-particle VMC study. For a detailed
description, we refer to the text and Appendix
B.\label{fig:compare_E} }
\end{figure}

In \fig{fig:compare_E}, we  compare the GA of the kinetic energy
$E^{(1)}$, and the expectation value $E^{(2)}$, of the remaining
terms in the $t-J$ model ($\langle \hat S_i \hat S_j \rangle$,
$\langle \hat n_i \hat n_j \rangle$, and the 3-site term) to those
from the grand canonical VMC. A good agreement between the VMC and
Gutzwiller results is seen, which confirms the validity of our
grand canonical Gutzwiller approximation (\eq{grandC}).

%\begin{figure}[h]
%  \centering
%  \includegraphics*[width=0.45\textwidth]{compare_Gutz.eps}
%  \caption{Total energy per site of the $t-J$ model
%  (all above mentioned terms) in units of $t$ ($t^2/U=1/20$) at particle
%  density $n=0.95$. The results for the Gutzwiller
%  approximations, which correspond to the grand canonical VMC (dashed) and the fixed-particle
%  VMC (solid), are illustrated as a function of the order parameter $\Delta_0$
%  for a d-wave.
%  \label{fig:compare_Gutz}}
%\end{figure}

In \fig{compare_VMC} and \fig{fig:compare_E}, we also show
Gutzwiller approximations for the fixed particle number VMC
\cite{gros_88}. Clearly, canonical and grand canonical approaches yield different energies
(as do the corresponding VMC studies ). We emphasize this is
because of the projection operator $P$, which changes the particle number
in a grand canonical scheme. For these two methods to yield the
same results, a fugacity corrected wave function must be used when
working in a grand canonical ensemble. Hence, all previous
speculations about the coincidence of these two VMC schemes in the
thermodynamic limit
have to be reformulated carefully.
%%%%%%%%%%%%%%%%%%%%%%%%%%%%%%%%%%%%%%%%%%%%%%%%%%%%%%%%%%%%
%%%%%%%%%%%%%%%%%%%%%%%%%%%%%%%%%%%%%%%%%%%%%%%%%%%%%%%%%
\section{Summary}

In this paper, we considered the effects of Gutzwiller projection on a state which
does not have fixed particle number. We showed that it is necessary to include
a fugacity factor when invoking the Gutzwiller approximation for such states.
The effects of projecting a number non-conserving BCS state were studied by
examining
the relation between particle number before and
after projection. We obtained an analytical expression,
\eq{p_after2}, and compared to variational
Monte Carlo data (\fig{fig:compare_Yoko}). We discussed the
discrepancies in the VMC results for projected BCS wave functions obtained in the 
canonical and grand canonical schemes, and presented a resolution.
In conclusion,
we have clarified several subtle properties of the Gutzwiller
projection operator $P$ acting on a BCS state, and hope that these results lead to a better
understanding of the Gutzwiller approximation in the grand canonical scheme.

\bigskip
We thank P.~W.~Anderson, N.~P.~Ong, and H.~Yokoyama for several
discussions. N.~F. was supported by the Deutsche
Forschungsgemeinschaft. V.~N.~M. acknowledges partial financial
support from The City University of New York, PSC-CUNY Research
Award Program.

%%%%%%%%%%%%%%%%%%%%%%%%%%%%%%%%%%%%%%%%%%%%%%%%%%%%%%%%%%%%
\begin{appendix}

\section{Saddle point approximation
to mean particle number and number fluctuations
in projected BCS wave functions}

In Sec. III, we used the fugacity factor to derive \eq{p_after2}.
Here, we present an alternative approach by a
saddle point approximation to discuss the effects of
projection on the mean particle number of a BCS state. This
approach also describes the particle number fluctuations after
projection.

The particle number distribution for an unprojected BCS
wave function $ \rho^{(0)}_N$ can be written as
\begin{equation}
\left. \rho^{(0)}_{N}=\frac{2}{N_\sigma !} \left(\frac{\rm d}{\rm
    d\lambda}\right)^{N_\sigma}
\prod_k \left(|u_k|^2+|v_k|^2\lambda\right)\right|_{\lambda\rightarrow0},\label{exact_rho}
\end{equation}
where ${N_\sigma}=N/2$ is the number of electron pairs. This
relation can be checked by expanding the product in the wave
function,
\begin{equation}
|\Psi_{\rm BCS} \rangle = \prod_k \left(u_k+v_k
c^\dagger_{k,\uparrow} c^\dagger_{-k,\downarrow} \right)|0\rangle
\nonumber \ ,
\end{equation}
and considering contribution to $\langle \Psi_{\rm BCS} |\Psi_{\rm
BCS} \rangle$ from each term. In Sec. III, we showed that the particle
number distribution of a projected wave function $\rho_N$ is
related to the unprojected distribution $ \rho^{(0)}_N$ by
\begin{equation}
 \rho_N =  g_N \ \rho^{(0)}_N \ , \label{rho_after_a1}
\end{equation}
where,
\begin{align}
g_N\approx \, C\,\frac{(\left (L-N/2)! \right)^2}{L!\,(L-N)!}\,
\nonumber.
\end{align}
Particle number and number fluctuations of the projected wave function
can be derived from \eq{rho_after_a1} and \eq{exact_rho}, upon invoking a saddle
point approximation.
We define a generating function $\Xi_\lambda$,
\begin{align}
 \Xi_\lambda \equiv & ~2 \prod_k \left(|u_k|^2+|v_k|^2\lambda\right) \nonumber \\
=&\sum_{{N_\sigma}=0}^L \lambda^{N_\sigma} \rho^{(0)}_{2 N_\sigma}
\ . \label{def_Xi}
\end{align}
We invert \eq{def_Xi} using a contour integral
on the complex $\lambda$-plane along a circle around $\lambda=0$:
\begin{align}
\rho^{(0)}_{N}=\frac{1}{2\pi i}\oint
\frac{\Xi_\lambda}{\lambda^{{N_\sigma}+1}} d\lambda\ .
\label{contour_unp1}
\end{align}
Note that, in the integrand, only $\rho^{(0)}_{2
N_\sigma}/\lambda=\rho^{(0)}_{N}/\lambda$ gives a finite value.
The others powers of $\lambda$ vanishes. Multiplying by $g_N$
gives
\begin{align}
\rho_{N} \approx \frac{1}{2\pi i}\oint g_N
\frac{\Xi_\lambda}{\lambda^{{N_\sigma}+1}} d\lambda
\label{contour_p1} \ .
\end{align}
\eq{contour_p1} can be written as
\begin{align}
\rho_{N} \approx  \frac{1}{2\pi i} \oint d\lambda \; e^{f(\lambda,N)}\ ,
\label{contour_p2}
\end{align}
where
\begin{align}
  f(\lambda,N)=\log \Xi_\lambda - {(\frac N 2+1)} \log \lambda \ +
  \log g_N.
\end{align}
Using Stirling's formula,
\begin{align}
\log g_N \approx& \ 2 \ {(L-\frac N 2)} \ \log {(L-\frac N 2)}
 -{(L-N)} \ \log {(L-N)}  \nonumber \\
-&L \ \log L\  + {\log C}
\end{align}
The saddle point ($\bar n = \frac {\bar N} L$, $\bar \lambda$) of
\eq{contour_p2} is determined by
\begin{align}
\frac{\partial f}{ \partial\lambda}=& \
\frac{ \partial \log \Xi_\lambda}{
  \partial\lambda}-\frac{\frac N 2+1}{\lambda}\nonumber \\
=& \ \frac{ \partial}{ \partial\lambda}
 \sum_k \log \left(|u_k|^2+|v_k|^2\lambda\right)-\frac{{\frac N 2}+1}{\lambda}
\nonumber \\
=&  \ \sum_k \frac{|v_k|^2}{|u_k|^2+|v_k|^2\lambda}-\frac{{\frac N 2}+1}{\lambda}
\nonumber \\
\equiv  & \ 0 \ ,
\end{align}
for $N \gg 1$,
\begin{align}
{\bar n} \approx& \ 2 \ \frac{\frac {\bar N} 2 +1}{L} \nonumber \\
=&  \ 2 \ \frac 1 L \sum_k \frac{\bar\lambda
|v_k|^2}{|u_k|^2+\bar\lambda
  |v_k|^2} \label{cond1_p}\ ,
\end{align}
and
\begin{align}
\frac{\partial f}{\partial N}= & \ - \frac 1 2 \log\lambda - \log (L-\frac N
 2) + \log (L-N) \nonumber \\
 \equiv & \ 0 \ ,
\end{align}
\textit{i.e.},
\begin{align}
 \bar\lambda =& \left( \frac{L-\bar N}{L-\frac {\bar N} 2} \right)^2 \nonumber \\
=&g_t^2 . \label{cond2_p}
\end{align}
\eq{cond1_p} and \eq{cond2_p} lead to \eq{p_after2},
\begin{align}
n_{\rm after}=\bar n = \frac {{\bar N}} L = 2\, \frac 1 L \,\sum_k \frac{g_t^2
|{v}_k|^2} {|{u}_k|^2 +g_t^2 \nonumber |{v}_k|^2}~,
\end{align}
for the average particle density of a projected BCS wave function.
Without the factor $g_N$ this calculation would give the well known result
for an unprojected BCS wave function,
\begin{align}
n_{\rm before}=\bar n &= \ 2 \ \frac 1 L \sum_k |v_k|^2 \ . \nonumber
\end{align}

\begin{figure}[t]
  \centering
 \includegraphics*[width=0.45\textwidth]{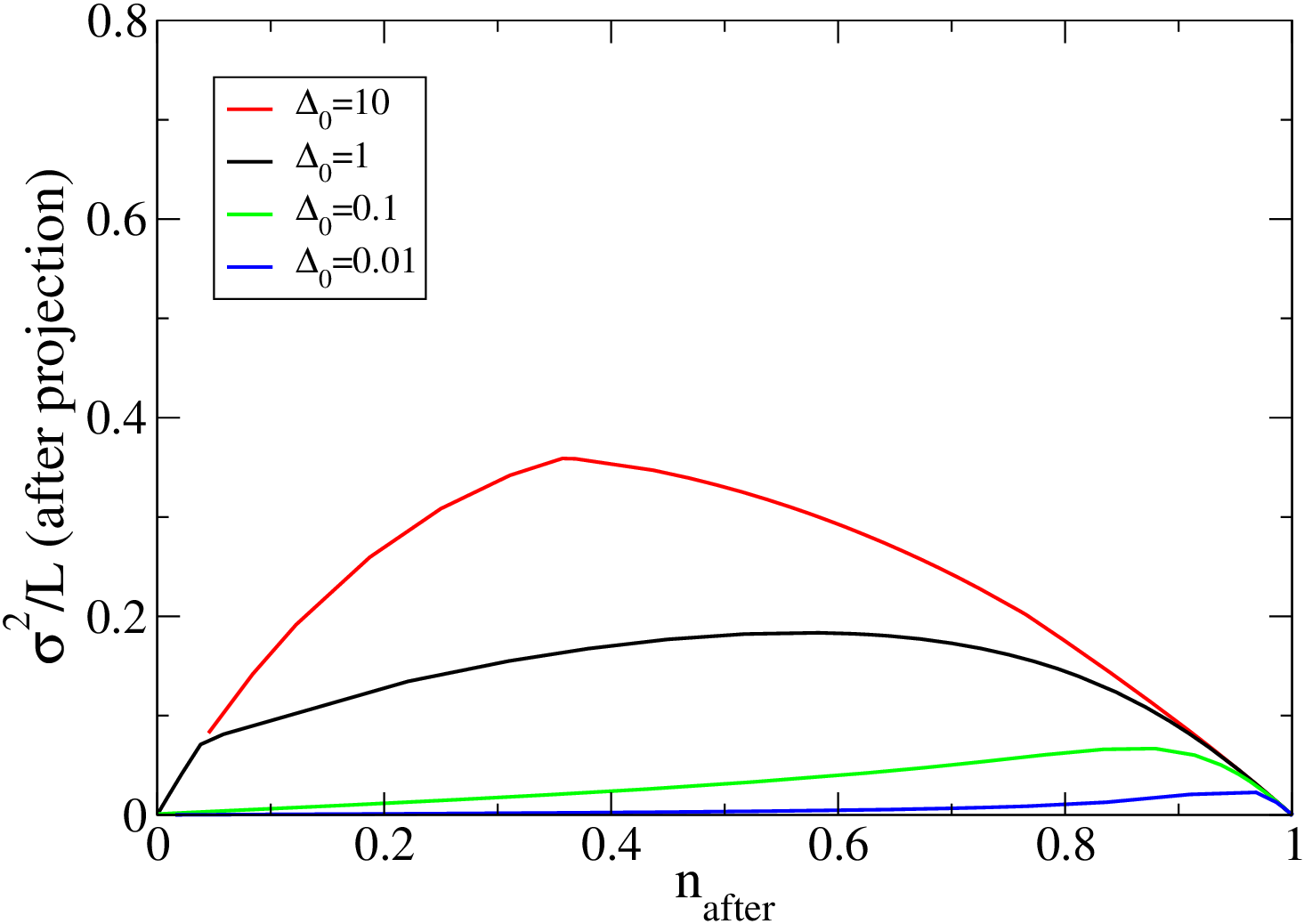}
 \includegraphics*[width=0.45\textwidth]{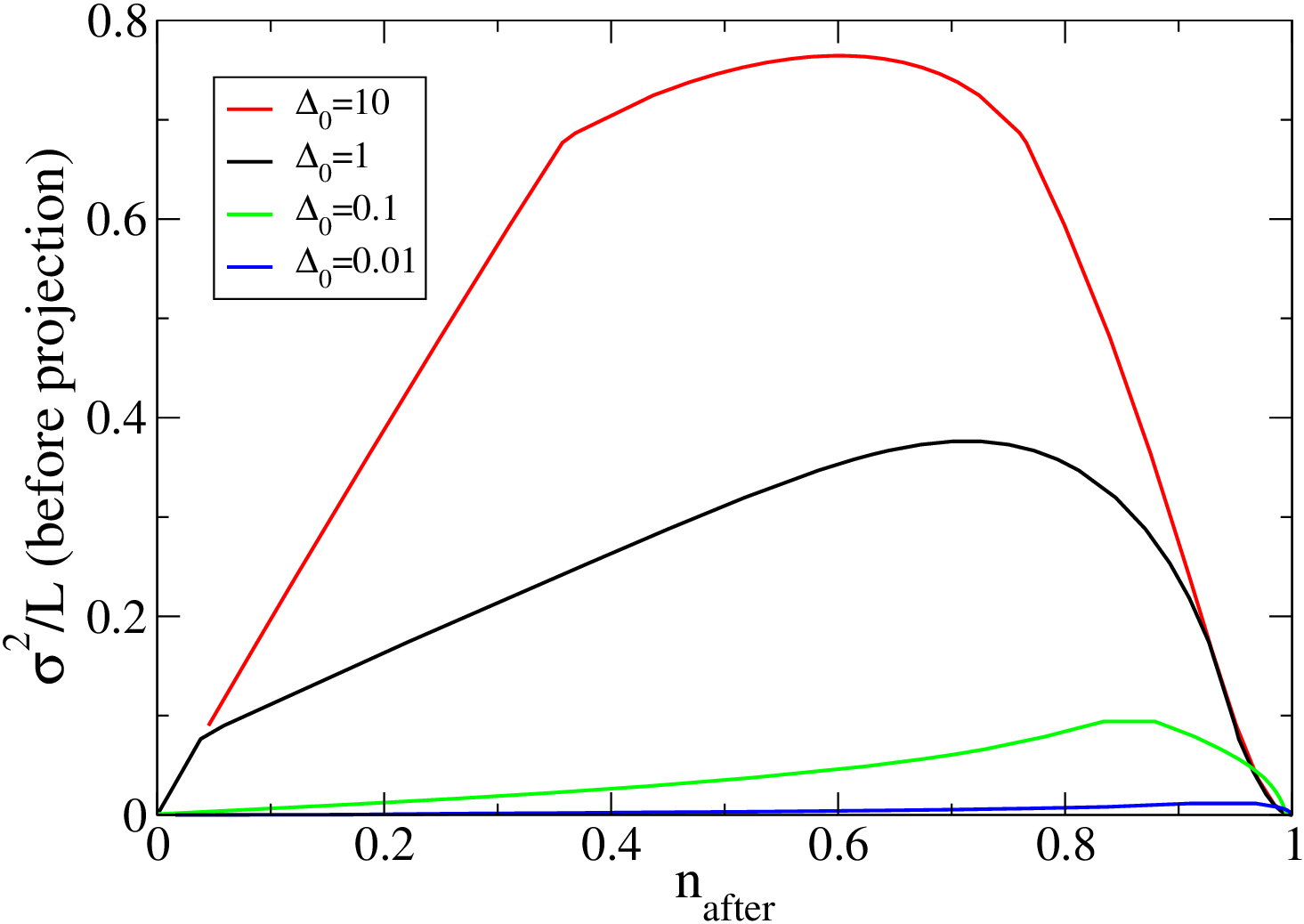}
 \caption{\label{fig:fluc}%
 (Color online)
 The fluctuation $\sigma_{\rm
 after}^2$/$\sigma_{\rm before}^2$ after/before projection
 for different values of the variational parameter
 $\Delta_0=0.01/0.1/1/10$, as a function of the particle
 density after projection $n_{\rm after}$. 
 \label{fig:all_na}
}
\end{figure}

To calculate the particle number fluctuations of the projected
wave function, we need to expand $f(\lambda,N)$ up to second order
in $N$ and $\lambda$ around the saddle point. Then,
integration over $\lambda$ in \eq{contour_p2} approximates the
particle number distribution $ \rho$ by a gaussian distribution,
yielding an expression for number fluctuations. With
\begin{align}
f_{\lambda\lambda}(\lambda,N)\equiv& \ \frac{\partial^2
f}{\partial\lambda^2}
\nonumber \\
=& \ -\ \sum_k \frac{|v_k|^4}{(|u_k|^2+|v_k|^2\lambda)^2}
+\frac{{\frac N 2 +1}}{\lambda^2} \ , \nonumber \\
f_{\lambda N}(\lambda,N)\equiv& \ \frac{\partial^2
f}{\partial\lambda\partial N}
=\ -\frac{1}{2 \lambda} \ , \nonumber \\
f_{NN}(\lambda,N)\equiv& \ \frac{\partial^2 f}{{\partial N}^2} =
\frac 1 2 \frac{1}{L-{\frac N 2}}-\frac{1}{L-{N}} \ , \nonumber
\end{align}
the second order expansion can be written as
\begin{align}
 f(\lambda,N)-f(\bar\lambda,\bar N)
\approx &f_{\lambda\lambda}(\bar\lambda,\bar N)
\frac{(\lambda-\bar\lambda)^2}{2} \nonumber \\
+&f_{\lambda N}(\bar\lambda,\bar N )
(\lambda-\bar\lambda)(N-\bar N) \nonumber \\
+&f_{NN}(\bar\lambda,\bar N) \frac{(N-\bar N)^2}{2} \ .
\end{align}
For this level of the saddle point approximation for $\lambda$,
the contour around $\bar\lambda$ for the integral in \eq
{contour_p2} must be taken so that
${f_{\lambda\lambda}(\bar\lambda,\bar N
)(\lambda-\bar\lambda)^2}{<0}$. Since
$f_{\lambda\lambda}(\bar\lambda,\bar N)>0$ and contribution only
near the saddle point is relevant, the path is taken from
$\bar\lambda-i\infty$ to $\bar\lambda+i\infty$. By variable
transformation $\lambda=\bar\lambda+i\lambda^\prime$, one can
perform a gaussian integral of $\lambda^\prime$. Then, we obtain a
a gaussian distribution for $ \rho_N$,
\begin{align}
& \rho_N \approx  \frac{1}{2\pi} \int_{-\infty}^\infty
d\lambda^\prime \; e^{f(\bar\lambda+i \lambda^\prime,N)} \nonumber
\\ &\approx \exp\left[ \left(f_{\bar N \bar N} -\frac{f_{\bar \lambda
\bar N}^2}{f_{\bar\lambda\bar \lambda}}\right) \frac{\left(N-\bar
N\right)^2}{2} \right] e^{f(\bar \lambda,{\bar N})} \ .
\label{gaussian}
\end{align}
The variance of $\bar N$ (average particle number of a projected
BCS wave function) can now be read off from \eq{gaussian}. We
get,
\begin{align}
\frac{\sigma_{\rm after}^2} L & = -\frac 1 L \left(f_{\bar N \bar N}
-\frac{f_{\bar \lambda \bar N}^2}{f_{\bar\lambda\bar
    \lambda}}\right)^{-1} \nonumber \\
&=2 \left[ \frac{1}{(1-\frac {\bar n} 2)(1- \bar n)}+
\frac{1}{\frac{2}{L}\sum_k \frac{\bar \lambda |u_k|^2 |v_k|^2
}{(|u_k|^2+|v_k|^2 \bar \lambda)^2}} \right]^{-1} ,
\end{align}
where we used \eq{cond1_p} in the second term. For $\bar \lambda$
we must insert $g_t^2$. For completeness, we mention that for the
unprojected wave function, \textit{i.e.} $g_N$ not included, this
approach yields the known result
\begin{align}
\frac{\sigma_{\rm before}^2} L = \ 4 \ \frac{1}{L} \sum_k |u_k|^2
|v_k|^2 \ .
\end{align}
The fluctuations $\frac{\sigma^2} L$ are illustrated in
\fig{fig:fluc} as a function of the particle density after
projection $n_{\rm after}\, (=\bar n)$ for unprojected ($|\,\Psi_{\rm
BCS}\rangle$) and projected ($P |\,\Psi_{\rm
BCS}\rangle$) BCS $d$-wave functions. As expected, the fluctuations vanish at half filling, since projection freezes the charge degrees of freedom entirely.

\section{Gutzwiller approximation for the $t-J$ Hamiltonian}
We summarize the Gutzwiller approximation for the so-called 3-site
terms in the $t-J$ model, that are included in the VMC study of
Yokoyama and Shiba \cite{yokoyama_88}.

The $t-J$ model can be derived from a large $U$ expansion of the
Hubbard model. The Hamiltonian is valid in the reduced
Hilbert space of no double occupied states, and is given by
\begin{equation}
H_{\rm eff}\ =\ T+H_{\rm eff}^{(2)} \label{tJmodel}
\end{equation}
where
\begin{equation}
T\ =\ -t\, \sum_{\langle i,j \rangle,\sigma} \left(
c_{i,\sigma}^\dagger c_{j,\sigma} + c_{j,\sigma}^\dagger
c_{i,\sigma} \right) \label{kin_tJ}
\end{equation}
and
\begin{align}
H_{\rm eff}^{(2)} =&\, J \, \sum_{\langle i,j \rangle}
{\bf S}_i \, {\bf S}_j
\,-\,\frac J 4 \sum_{\langle i,j \rangle} n_i n_j \nonumber \\
-&\frac J 4 \, \sum_{i,\tau \neq \tau^\prime, \sigma}
c_{i+\tau,\sigma}^\dagger c_{i,{-\sigma}}^\dagger c_{i,{-\sigma}}
c_{i+\tau^\prime,\sigma} \nonumber \\
+&\frac J 4 \, \sum_{i,\tau \neq \tau^\prime, \sigma}
c_{i+\tau,{-\sigma}}^\dagger c_{i,\sigma}^\dagger c_{i,{-\sigma}}
c_{i+\tau^\prime,\sigma} \ . \label{H2_tJ}
\end{align}
Here, $J=4 \frac {t^2} {U}$, ${\bf S}_i$ are the spin operators on site
$i$, and $n_i=n_{i,\uparrow}+n_{i,_\downarrow}$ with
$n_{i,\sigma}=c_{i,\sigma}^\dagger c_{i,\sigma}$. $\langle i,j
\rangle$ are pairs of n.n sites and $i+\tau$ denotes a n.n. site
of $i$.

We are interested in the energies $E^{(1)}$ and $E^{(2)}$
calculated in Ref. \onlinecite{yokoyama_88}:
\begin{align}
E^{(1)}&=\frac 1 L \frac{\langle
\Psi_{\rm BCS}|\,P~T~P\,|\Psi_{\rm BCS}\rangle}{\langle
\Psi_{\rm BCS}|\,P~P\,|\Psi_{\rm BCS}\rangle} \ , \nonumber \\
E^{(2)}&=\frac 1 L \frac{\langle
\Psi_{\rm BCS}|\,P~H^{(2)}_{\rm eff}~P\,|\Psi_{\rm BCS}\rangle} {\langle
\Psi_{\rm BCS}|\,P~P\,|\Psi_{\rm BCS}\rangle} \label{E1_E2_1}\ .
\end{align}
We invoke the Gutzwiller approximation. The renormalization
factors, $g_t$ for kinetic energy (\eq{kin_tJ}) and $g_S$ for spin
exchange (first term in \eq{H2_tJ}), are given in \eq{g_t_s}. The
second term of \eq{H2_tJ}, $n_i n_j$, is not renormalized. The
approximation for the 3-site terms (3rd and 4th term of
\eq{H2_tJ}) is done as follows ($| \psi \rangle = P | \psi_0
\rangle$):
\begin{align}
&\frac{\langle \Psi | c_{i+\tau,\uparrow}^\dagger
c_{i,{\downarrow}}^\dagger c_{i,{\downarrow}}
c_{i+\tau^\prime,\uparrow}|\Psi \rangle}{\langle \Psi | \Psi \rangle} \nonumber\\
&\qquad = \, \frac{\langle \Psi | c_{i+\tau,\uparrow}^\dagger
n_{i,\downarrow} (1-n_{i,{\uparrow}})
c_{i+\tau^\prime,\uparrow}|\Psi \rangle}{\langle \Psi | \Psi \rangle} \nonumber\\
&\qquad = g_{3} \, \frac{\langle \Psi_0 |
c_{i+\tau,\uparrow}^\dagger n_{i,\downarrow} (1-n_{i,{\uparrow}})
c_{i+\tau^\prime,\uparrow}|\Psi_0 \rangle}{\langle \Psi_0 | \Psi_0
\rangle}
\label{Gutz_3a} \ ,\\
&\frac{\langle \Psi| c_{i+\tau,\downarrow}^\dagger
c_{i,\uparrow}^\dagger c_{i,\downarrow}
c_{i+\tau^\prime,\uparrow}|\Psi \rangle}{\langle \Psi | \Psi \rangle} \nonumber\\
&\qquad = g_{3} \, \frac{\langle \Psi_0 |
c_{i+\tau,\downarrow}^\dagger c^\dagger_{i,\uparrow}
c_{i,\downarrow} c_{i+\tau^\prime,\uparrow}|\Psi_0
\rangle}{\langle \Psi_0 | \Psi_0 \rangle} \label{Gutz_3b}
\end{align}
The renormalization factor $g_{3}$ is derived by considering the
number of terms that contribute to the projected and the
unprojected side respectively. The projected side (lhs) contributes only
if (i) site $i+\tau$ is unoccupied, \textit{i.e.}, probability
$(1-n)$, (ii) site $i$ is singly occupied by a
$\downarrow$-electron, \textit{i.e.}, probability $n_\downarrow$,
and (iii) site $i+\tau^\prime$ is singly occupied by an
$\uparrow$-electron, \textit{i.e.}, probability $n_\uparrow$. On
the other hand, the unprojected side (rhs) in \eq{Gutz_3a}/\eq{Gutz_3b}
contributes only if (i) site $i+\tau$ is not occupied by an
$\uparrow$-electron/$\downarrow$-electron, \textit{i.e.},
probability $(1-n_{\uparrow})$ / $(1-n_{\downarrow})$ , (ii) site $i$
is singly occupied by a $\downarrow$-electron, \textit{i.e.},
probability $n_{\downarrow}(1-n_{\uparrow})$, and (iii) site
$i+\tau^\prime$ must have an $\uparrow$-electron, \textit{i.e.},
probability $n_\uparrow$. These probabilities yield the Gutzwiller
factor (ratio of contributions from projected and unprojected
states),
\begin{align}
 g_{3}=\frac{(1-n) n_\sigma n_\sigma}
{(1-n_{\sigma})n_{\sigma}(1-n_{\sigma})n_{\sigma}}=\frac{1-n}{(1-n_\sigma)^2}
\label{g_3s} \, ,
\end{align}
where we assumed $n_\uparrow=n_\downarrow=n_\sigma$.

We can now write down the renormalized $t-J$ Hamiltonian
$H^\prime_{\rm eff}$,
\begin{equation}
H^\prime_{\rm eff}\ =\ T^\prime+{H'}_{\rm eff}^{(2)}~, \label{tJmodel_renorm}
\end{equation}
where,
\begin{equation}
T^\prime\ =\ -g_t\,t\, \sum_{\langle i,j \rangle,\sigma} \left(
c_{i,\sigma}^\dagger c_{j,\sigma} + c_{j,\sigma}^\dagger
c_{i,\sigma} \right)~, \label{kin_tJ_renorm}
\end{equation}
\begin{align}
{H'}_{\rm eff}^{(2)} =&\, g_S\,J \, \sum_{\langle i,j \rangle}
{\bf S}_i \, {\bf S}_j
\,-\,\frac J 4 \sum_{\langle i,j \rangle} n_i n_j \nonumber \\
-&g_{3}\, \frac J 4 \, \sum_{i,\tau \neq \tau^\prime, \sigma}
c_{i+\tau,\sigma}^\dagger n_{i,{-\sigma}} (1-n_{i,{\sigma}})
c_{i+\tau^\prime,\sigma} \nonumber \\
+&g_{3}\, \frac J 4 \, \sum_{i,\tau \neq \tau^\prime, \sigma}
c_{i+\tau,{-\sigma}}^\dagger c_{i,\sigma}^\dagger c_{i,{-\sigma}}
c_{i+\tau^\prime,\sigma} \ . \label{H2_tJ_renorm}
\end{align}
By using $T^\prime$ and ${H'}_{\rm eff}^{(2)}$, \eq{E1_E2_1}
($E^{(1)}$ and $E^{(2)}$) can be calculated using unprojected
wave functions,
\begin{align}
E^{(1)}&=\frac 1 L \frac{\langle
\Psi_{\rm BCS}|\,T'\,|\Psi_{\rm BCS}\rangle}{\langle
\Psi_{\rm BCS}|\Psi_{\rm BCS}\rangle} \ ,\nonumber \\
E^{(2)}&=\frac 1 L \frac{\langle
\Psi_{\rm BCS}|\,{H'}^{(2)}_{\rm eff}\,|\Psi_{\rm BCS}\rangle} {\langle
\Psi_{\rm BCS}|\Psi_{\rm BCS}\rangle} \label{E1_E2_2}\ .
\end{align}

Evaluating \eq{E1_E2_2} by Wick's decomposition for a
$d$-wave BCS state,
$$
\frac {E^{(1)}} t\ =\ -2 g_t (\xi_x+\xi_y) \ ,
$$
\begin{align}
\frac {E^{(2)}}{0.25 J} =& -(3 g_s -1)
(\xi_x^2+\xi_y^2)/2 \label{rest_terms} \\
&-(3 g_s+1) (|\tilde \Delta_x|^2 + |\tilde \Delta_y|^2)/2
\,-\,2 n^2 \nonumber \\
&-g_3\, n \,(1-n_\sigma)\,
(\xi_{2x}+\xi_{2y}+2 \xi_{x-y} + 2 \xi_{x+y}) \nonumber \\
&-g_3\, (1+n_\sigma)\,(\xi_x^2+\xi_y^2+4 \xi_x \xi_y) \nonumber \\
&- g_3\, (2-n_\sigma)\,( |\tilde \Delta_x|^2 +
|\tilde \Delta_y|^2 +4 \Re(\tilde \Delta_x
\tilde \Delta_{y}^*)) \nonumber ~ ,
\end{align}
where we defined ($\tau,\tau^{\prime} = x,y$)
\begin{align}
\xi_{\tau}&=\sum_\sigma \langle c^\dagger_{i,\sigma} c_{i+\tau,\sigma} \rangle=
\frac 1 L \sum_k 2 \cos(k_{\tau}) |v_k|^2 \ ,\nonumber \\
\xi_{\tau \pm \tau^{\prime}}&=
\sum_\sigma \langle c^\dagger_{i+\tau,\sigma} c_{i\pm \tau^{\prime},\sigma}  \rangle=
\frac 1 L \sum_k 2 \cos(k_{\tau} \pm k_{\tau^{\prime}}) |v_k|^2\nonumber \ , \\
\tilde
\Delta_{\tau}&=\langle c^\dagger_{i,\uparrow} c^\dagger_{i+\tau,\downarrow} -
c^\dagger_{i,\downarrow} c^\dagger_{i+\tau,\uparrow}
 \rangle=
 \frac 1 L \sum_k 2 \cos(k_{\tau}) v_k
u_{k}^*\nonumber \ .
\end{align}
The last three rows in \eq{rest_terms} correspond to the 3-site
terms of the $t-J$ model and are renormalized by the Gutzwiller
factor $g_3$. Here, it's important to note that
the order parameter $\tilde \Delta_\tau$ is related, but not identical,
to the previous introduced variational parameter $\Delta_0$.
\end{appendix}

%%%%%%%%%%%%%%%%%%%%%%%%%%%%%%%%%%%%%%%%%%%%%%%%%%%%%%%%%

%%%%%%%%%%%%%%%%%%%%%%%%%%%%%%%%%%%%%%%%%%%%%%%%%%%%%%%%%%%%

\end{document}